\documentclass{article}
\usepackage[utf8]{inputenc}
\usepackage[a4paper, total={6.8in, 10in}]{geometry}
\usepackage{authblk}
\usepackage{graphicx}
\graphicspath{{figures/}}
\usepackage[bottom]{footmisc}
\usepackage{caption}

\title{Team Cogitat at NeurIPS 2021:\\Benchmarks for EEG Transfer Learning Competition}

\author[1, 2]{Stylianos Bakas\footnote{please address correspondence to stelios@cogitat.io}}
\author[1, 2]{Siegfried Ludwig}
\author[1, 2]{Konstantinos Barmpas}
\author[1, 2]{Mehdi Bahri}
\author[1, 2]{Yannis Panagakis}
\author[1, 2]{Nikolaos Laskaris}
\author[1, 2]{Dimitrios A. Adamos}
\author[1, 2]{Stefanos Zafeiriou}

\affil[1]{Cogitat Ltd., London, U.K.}
\affil[2]{Department of Computing, Imperial College London, U.K.} 

\date{October 2021}

\begin{document}

\maketitle

\section{Introduction}

Building subject-independent deep learning models for EEG decoding faces the challenge of strong covariate-shift across different datasets, subjects and recording sessions. Our approach to address this difficulty is to explicitly align feature distributions at various layers of the deep learning model, using both simple statistical techniques as well as trainable methods with more representational capacity. This follows in a similar vein as covariance-based  alignment methods \cite{he2019transfer}, often used in a Riemannian manifold context \cite{zanini2017transfer}.

The methodology proposed herein won first place in the 2021 Benchmarks in EEG Transfer Learning (BEETL) competition\footnote{https://beetl.ai/introduction}, hosted at the NeurIPS conference\footnote{https://neurips.cc/Conferences/2021/CompetitionTrack}. The first task of the competition consisted of sleep stage classification, which required the transfer of models trained on younger subjects to perform inference on multiple subjects of older age groups without personalized calibration data, requiring subject-independent models. The second task required to transfer models trained on the subjects of one or more source motor imagery datasets to perform inference on two target datasets, providing a small set of personalized calibration data for multiple test subjects.

\section{Methodology}

The decoding model used in both tasks is a convolutional neural network architecture specifically designed for EEG decoding. We introduce minor modifications to the EEGInception architecture \cite{santamaria2020eeg} to reduce the number of trainable parameters while preserving the representational power of the model. Specifically, we introduce dilations to the larger convolutional kernels in the inception layers with accompanying average-pooling functions on the respective filter inputs, as well as separable convolution arrangements in later model stages.

\begin{figure*}[!b]
    \centering
    \includegraphics[width=0.85\textwidth]{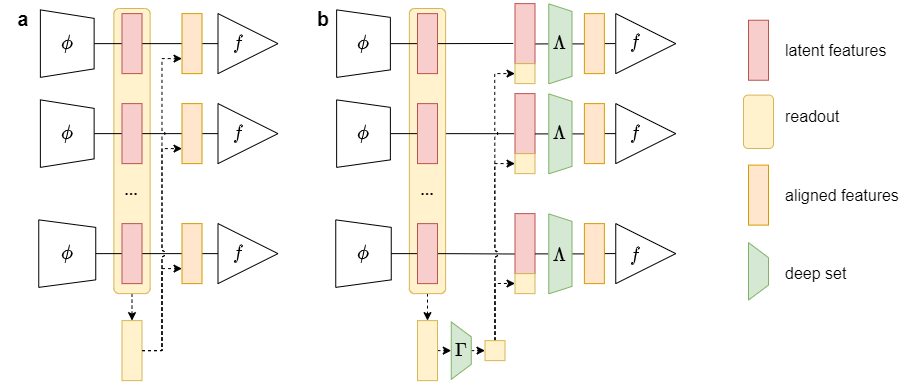}
    \caption{Schematic representation of feature alignment between sessions or subjects. The model input is transformed by a feature extractor $\phi$, followed by feature alignment. After alignment, the latent features are used in the classification function $f$ to make a prediction. a) Statistical feature alignment uses statistics of latent features across trials of a given subject to standardize each feature distribution. b) A modified deep set uses a trainable function $\Gamma$ to transform the readout across trials and use this readout to update each feature vector with a trainable function $\Lambda$.}
    \label{fig:alignment}
\end{figure*}

Our models are trained on source and calibration data simultaneously, while oversampling the calibration set to give it equal weight during training. To estimate model performance we perform cross-validation, leaving out part of the calibration set for each fold. During inference on the test data, model predictions are ensembled over cross-validation folds, using the mean of output logits. Performance is measured in unweighted average recall (UAR).

To perform feature alignment for each subject (or session, where applicable), models are trained using chunks of subject-wise trials. Each training minibatch therefore consists of multiple subjects, with multiple trials from each included subject. During inference, we predict on all trials of a given subject/session simultaneously.

All random seeds are fixed and deterministic algorithms are used, making our experiments fully reproducible, given the same compute hardware.

\subsection{Sleep Task}

For the sleep task, the EEGInception model uses 8 temporal filters per inception branch and 4 spatial filters per resulting filter channel. We use increased dilations for the filters of the first inception module to increase the receptive field and focus on lower frequencies. Further, the number of filter channels is kept constant throughout the later stages of the model, the second pooling kernel is increased in size and global average pooling is used in the final layer to collapse the long temporal nature of each trial. As recordings of the sleep task are not time-locked, any temporal resolution in the linear classifier does not contribute to classification performance.

On the sleep task our model uses statistical feature alignment at every layer, including the model input, in order to align feature distributions of different subjects and sleep sessions. Specifically, the mean and standard deviation of each channel is computed across subject trials and the temporal dimension, and the feature distributions across trials are then standardized accordingly (see figure \ref{fig:alignment}, panel a). In resemblance to conventional batch normalization, the standardization is followed by a trainable weight and bias for each channel.

As the nature of the task requires strong subject-independent models, we include phase 1 ("leaderboard") data in the training set.

The sleep recordings show a strong class imbalance between different sleep stages. In order to give equal importance to each class, as the unweighted average recall metric of the competition requires, we use class weights in the cross-entropy loss function according to the distribution of classes on the oversampled training set.

The model is trained for 15 epochs with the Adam optimizer using a learning rate of $1\mathrm{e}{-3}$, weight decay of $1\mathrm{e}{-3}$ and dropout $0.25$. Cross-validation on the calibration set of five subjects was performed in a leave-one-subject-out fashion for two times, resulting in 10 total folds. The GPU hardware used was a Nvidia TITAN Xp.

\subsection{Motor Imagery Task}

The EEGInception model employed on the motor imagery task uses 8 temporal filters per inception branch and 2 spatial filters per resulting filter channel. We increased the kernel size of the last pooling layer to reduce the dimensionality of the final representation.

As the nature of the motor imagery task requires training on multiple distinct datasets, we use a shared feature extraction stage with separate classifiers for each dataset \cite{huang2013cross}\cite{wei2021inter}. The model is trained on all datasets simultaneously, indexing for the corresponding model output on each trial.

We perform statistical feature alignment per input channel (see figure \ref{fig:alignment}, panel a). This corresponds to performing input signal standardization across trials of each subject and thereby aligns the distribution of signal powers among different subjects. Specifically, we compute the mean and standard deviation across subject trials and the temporal dimension for each input channel and use these statistics to standardize the corresponding signals.

Our deep set feature alignment method is used during the later feature extraction stage as well as on the final feature representation \cite{zaheer2017deep}. For this, we perform a mean readout across subject trials for each feature. This is closely related to the simple statistical mean estimation of the previously mentioned statistical alignment. The resulting readout vector is transformed by a non-linear function while reducing its dimensionality and then concatenated to the feature vector of each trial. The resulting feature vector is transformed non-linearly and then used for classification (see figure \ref{fig:alignment}, panel b).

Since the motor imagery task provides calibration data for every test subject, the trained model does not have strict requirements for subject-independence. We therefore exclude phase 1 data ("leaderboard") to train each model head more specifically for the test subjects of each dataset. As additional training data we use PhysioNet motor imagery source data, including left fist, right fist, feet and true resting state trials (runs 1 and 2). We dropped six subjects due to various irregularities, including different sampling rates and number of recorded trials. In order to homogenize the three resulting datasets (PhysioNet, competition A and competition B), we only use the 30 common electrodes, resample all datasets to 160Hz and apply a 2Hz Butterworth highpass filter as well as Notch filters at both 50 and 60Hz. This is followed by common average referencing for the trials of each dataset.

We use a cross-entropy loss with label smoothing \cite{muller2019does}, which may reduce fold-wise performance but improves the results of ensembling. Although model performance is rated on three classes in the context of the competition, combining feet and resting state trials, we train our model on all four classes and combine predictions later. While trial numbers are balanced between these four classes, this is no longer true when combining feet and resting state. We therefore use class weights in the loss accordingly to give equal importance to the three logical classes (half importance for feet and resting state).

The model is trained for 200 epochs with the Adam optimizer using a learning rate of $5\mathrm{e}{-4}$, weight decay of $1\mathrm{e}{-3}$ and dropout $0.25$. Cross-validation on the calibration set was performed over 10 unstratified folds. The GPU hardware used was a Nvidia GeForce RTX 2080 Ti.

\section{Results and Discussion}

\begin{table}[]
\centering
\captionsetup[]{skip=5pt}
\caption{Results of baseline and alignment transfer learning models for the sleep and motor imagery tasks. The performance is measured as unweighted average recall in percent, and given as mean and standard deviation across folds.}
\label{tab:results}
\begin{tabular}{lll}
\hline
                            & Sleep             & Motor Imagery     \\ \hline
Baseline (validation)       & 56.87$\pm$7.70    & 63.72$\pm$4.76    \\
Alignment (validation)      & 67.70$\pm$4.20    & 72.44$\pm$5.30    \\ \hline
Alignment (test)            & 65.55             & 76.33             \\ \hline
\end{tabular}
\end{table}

On the six-class sleep stage classification task, our baseline model without statistical or deep set alignment achieved a mean UAR of 56.89\% during cross-validation (see table \ref{tab:results}). Note that this baseline includes training and oversampling on the calibration set. Adding statistical alignment to this model increased the cross-validation score to 67.70\%. This model is our entry to the competition and reached 65.55\% on the final test set after ensembling.

The baseline model for the four-class motor imagery task reached 63.72\%, including training and oversampling on the calibration set, as well as separate classifiers for each of the three datasets. After adding statistical input alignment and the deep set alignment method at later stages, this model achieves 72.44\% during cross-validation. As our entry to the competition, this model achieves 76.33\% on the final test set after ensembling.

The statistical and deep set alignment methods used in our models display very strong improvements in inference performance, both on the subject-independent sleep task as well as the more personalized motor imagery task, while only requiring a simple statistical estimation of feature distributions during inference. This effect is on top of more conventional transfer learning we employed by training on the domain-specific calibration set.

\bibliographystyle{unsrt}
\bibliography{references}

\end{document}